\documentclass{article}

\usepackage{arxiv}

\usepackage[utf8]{inputenc} 
\usepackage[T1]{fontenc}    
\usepackage{xurl} 
\usepackage{hyperref}       
\usepackage{url}            
\usepackage{booktabs}       
\usepackage{amsmath}
\usepackage{amsfonts}       
\usepackage{nicefrac}       
\usepackage{microtype}      
\usepackage{graphicx}
\usepackage{natbib}
\usepackage{doi}
\usepackage{subcaption}
\usepackage{float}

\usepackage[english]{babel}
\usepackage[autostyle]{csquotes}
\MakeOuterQuote{"}

\title{Exploring Climate-Related Anxiety Through Social Media Content\thanks{For more information, visit our \href{https://www.sprout-climate.org/climatelens/?ref=sprout-climate.org}{page}}}

\author{
    Karim El-Sharkawy \\
    Sprout Climate Association \\
    Troy, MI, US \\
    \texttt{karimelsharkawy2002@gmail.com} \\
\And 
    Zainab Rehman \\
    Sprout Climate Association \\ 
    Toronto, ON, CA \\ 
    \texttt{zainabbrehman@gmail.com} \\ 
\And
    Vikrant Kumar \\
    Sprout Climate Association \\
    Ajax, ON, CA \\
    \texttt{kumar.vikrant@outlook.com}
\And
    Ardavan Shahrabi \\
    Sprout Climate Association \\
    Toronto, ON, CA \\
    \texttt{ardavan.shahrabi@gmail.com} \\
\And
    Bobbie Williams \\
    Sprout Climate Association \\
    Toronto, ON, CA \\
    \texttt{williams.bobbie.g@gmail.com} \\
}

\renewcommand{\shorttitle}{Climate Anxiety in Social Media}

\hypersetup{
pdftitle={Climate Anxiety in Social Media: A Preprint},
pdfsubject={cs.SI, cs.CL},
pdfauthor={Karim El-Sharkawy, Zainab Rehman, Vikrant Kumar, Ardavan Shahrabi, Bobbie Williams},
pdfkeywords={climate anxiety, natural language processing (NLP), topic modelling, emotion classification, social media analysis},
}

\begin{document}
\maketitle

\begin{abstract}
	This study explores climate-related anxiety as expressed through social media discussions on Reddit. Using natural language processing techniques, we analyse large-scale textual data to identify recurring themes, emotional patterns, and how these evolve over time. Text data was preprocessed and analysed using BERTopic for topic modelling and a transformer-based model for emotion classification across 28 categories using the RoBERTa-base GoEmotions model. Results show that climate-related discourse is structured around a small number of core themes, primarily separating action-oriented discussions like advocacy and policy from informational and reflective content. Emotional analysis reveals that negative emotions such as fear and sadness are more prominent in posts, while comments often introduce a wider range of responses, including care, encouragement, and neutral reactions. These findings suggest that online climate discourse is shaped not only by the topics being discussed, but also by how users engage with and respond to one another. This work provides an initial framework for understanding climate-related anxiety through large-scale social media analysis and highlights opportunities for improving future models, expanding platform coverage, and incorporating youth-centred perspectives.
\end{abstract}


\section{Introduction}

Climate change is increasingly recognised not only as an environmental and political issue, but also as a psychological one. Climate anxiety and related forms of distress can involve fear, worry, sadness, guilt, helplessness, and uncertainty about the future \cite{dodds2021}. These responses are particularly relevant among younger people, who may face the consequences of climate change over much of their lives. A global survey of 10,000 young people aged 16--25 across ten countries found widespread concern about climate change and reported substantial levels of negative emotional responses among respondents \cite{hickman2021}.

Social media provides a complementary setting for examining these experiences as they are expressed in everyday communication. Unlike surveys and interviews, which depend on participants responding to structured questions, social media contains naturally occurring discussions in which users can express concerns, exchange information, and respond to one another. Social networks have also become an important part of contemporary climate-change communication, facilitating the dissemination of climate-related information, dialogue, and mobilization while enabling a wider range of actors to participate in climate discussions \cite{schäfer2025}.

The scale of social media data creates opportunities for computational approaches to examining these discussions. Natural language processing (NLP) can be used to identify recurring themes and patterns of emotional expression across large collections of text, while also allowing researchers to examine how these patterns vary across different types of interactions. Recent research has demonstrated the potential of social-media data for studying climate-anxiety discourse at scale, including the analysis of more than 177,000 Weibo posts to identify prominent themes and emotional patterns associated with climate anxiety \cite{tan2026}. Building on this emerging direction, this study examines climate-related discussions on Reddit using topic modelling and emotion classification. We identify recurring themes, examine patterns of emotional expression, investigate differences between original posts and comments, and explore how climate-related discussions evolve over time.

This paper presents an initial version (V1) of the ClimateLens framework, focusing on establishing a foundational analytical pipeline and identifying high-level patterns in climate-related discourse. The present analysis focuses exclusively on Reddit, while future iterations of the framework are intended to incorporate broader datasets, additional platforms, improved model validation, and more targeted analysis of youth perspectives.

The remainder of this paper is organised as follows. Section~\ref{literature review} reviews the existing literature. Section~\ref{methodology} describes the data preprocessing, topic modelling, temporal analysis, emotion classification, and visualisation methods used in the study. Section~\ref{results and discussion} presents and discusses the thematic and emotional patterns identified across the Reddit dataset. Section~\ref{limitations} discusses the limitations of the current framework and directions for future research. The paper concludes with a summary of the findings and broader implications of the ClimateLens framework.

\section{Literature Review}\label{literature review}

Climate anxiety or eco anxiety is defined as distress that shows up as a result of climate change and its impacts on how people navigate through their lives.  It could show up in forms as debilitating as panic attacks, loss of appetite, PTSD \cite{dodds2021} and as regular as in planning holidays, planning their finances, raising children.  It is even more apparent in younger people; a national survey with 1000 respondents in Canada, aged 16-25, reported that over a quarter of the respondents reported that climate change has a huge bearing on their mental health and 37\% said that climate change impacts how they carry out their regular activities \cite{galway2023}. These results are mirrored in other countries, too: in a study conducted in Nigeria, India, Australia, Brazil, Philippines, Finland, France, Portugal, the UK, and the US, where 1000 respondents per country in the same age group were asked questions about their feelings related to climate change \cite{hickman2021}. Of the total respondents, 75\% found the future frightening and more than half reported feeling negatively (sad, guilty, worried). In both the studies, there is a sense of disillusionment when it comes to government response: how governments have tackled climate change so far is more in line with “betrayal than reassurance”.

Climate change is a collective challenge that is experienced and addressed at the individual level, yet its causes and consequences transcend individual action and ultimately require collective efforts to address.
That, in addition to people’s lack of trust in their governments with regard to tackling climate change, the younger population turns to social media—which is considered freer than traditional media \cite{poushter2025} is not surprising.
Distress and anxiety of all sorts can be isolating and having a space where one can communicate with like-minded people is a source of relief. That's where social media comes in; it allows young people to access resources on how to engage in environmental behavior, as well as connecting with peers like-minded \cite{crandon2022}, which could potentially empower them to manage their climate anxiety.

Given how complex and nuanced these discussions are, computational methods allow the analysis of themes and the understanding of emotional dimensions by employing various methods. For one, Sentiment analysis has been used to categorise tonality on the basis of whether it is positive, neutral, or negative. For a more granular insight, emotion analysis dives deep into the text and helps uncover layers that sentiment analysis does not capture. These methods do not come without further complexities: both these analyses require methodological approaches that involve but aren’t limited to handling sarcasm, profanity, stopwords, using words from a different language \cite{zhang2022, birjali2023}.
Besides looking at emotional degrees within which the text operates, other methods include topic modelling using Latent Dirichlet Allocation (LDA). LDA posits that documents are created by drawing from a mixture of underlying (latent) topics, with each topic contributing to the words in the document \cite{murel2025}. For short social media texts, BERTopic, a topic modelling framework that uses transformer embeddings, might be more effective at finding semantic relationships. That said, the choice of method depends on semantic coherence, interpretability, and computational efficiency; all of which is even more enhanced when it comes to something as emotionally charged and nuanced as eco anxiety discussions on social media.

Existing research has examined the relationship between climate anxiety, social media, and emotional responses to climate change, while recent computational studies have begun to analyse climate-anxiety discourse at scale. However, there remains a need for approaches that jointly examine the thematic structure, fine-grained emotional expression, temporal evolution, and interaction dynamics of climate-related discussions across online communities. This paper addresses this gap by combining topic modelling and emotion classification to examine climate-related discourse on Reddit, with particular attention to differences between original posts and community responses.

\section{Methodology}\label{methodology}

\subsection{Preprocessing and Cleaning}\label{preprocessing-and-cleaning}

\textbf{Filtering Climate-Relevant Content.} To ensure that analysed content was genuinely related to climate anxiety rather than general mental health discourse, a keyword-based filtering step was applied prior to preprocessing. A predefined list of 27 search terms was developed, combining direct climate-related phrases (`climate change', `global warming', `solastalgia', `ecological grief') with anxiety- and distress-related terms specifically tied to climate concerns (`eco-anxiety', `climate doom', `eco-paralysis', `helplessness', `collective guilt').

For each JSONL input file (distinguishing between submissions and comments via the presence of a \texttt{"body"} field), the filtering function extracted text from \texttt{"selftext"} or \texttt{"title"} for posts, or \texttt{"body"} for comments. Case-insensitive matching was applied to check whether \emph{any} keyword appeared in the text. Only records containing at least one keyword were retained. This approach intentionally prioritises recall over precision, so some posts may mention climate change only peripherally but avoids the more serious risk of including purely general anxiety or depression content that lacks a climate framing. Matched records were written to CSV files with \texttt{subreddit}, filtered \texttt{body}, and \texttt{created\_utc} fields for downstream analysis.

The filtering step was applied before any text cleaning (e.g., stopword removal, tokenisation) to preserve original phrasing for accurate keyword matching. Empty or malformed JSON lines were skipped silently, and files with zero matches were reported but not written. This procedure ensures that subsequent topic modelling and emotion classification are grounded in climate-relevant discourse rather than general mental health discussions.

\textbf{Text Preprocessing.} Text preprocessing was performed to prepare Reddit data for downstream analysis. Raw text underwent multiple cleaning steps to remove noise while preserving semantically meaningful content. The resulting text was used for both topic modelling (semantic clustering) and emotion analysis (where preserving negations and modality improves classification accuracy).

URLs were removed using regular expressions targeting HTTP/HTTPS links and common web patterns (e.g., `www'). HTML entities were decoded and stripped. Residual URL fragments (e.g., `https', `co', `pic') were removed using additional regex filtering.

Tokenisation was performed using NLTK's \texttt{word\_tokenize}. A custom stopword list extended standard English stopwords with platform-specific and non-informative terms (e.g., `upvote', `subreddit'). Negations and modal verbs (e.g., `not', `no', `should', `could') were retained to preserve sentiment and contextual meaning.

Additional cleaning steps included removing consecutive duplicate tokens and filtering out documents with fewer than three words, as these lacked sufficient semantic content for reliable topic assignment.

\subsection{Topic modelling}\label{topic-modelling}

Topic modelling was conducted using BERTopic, selected for its ability to capture semantic relationships through transformer-based embeddings rather than relying solely on word co-occurrence (as in LDA). This makes it particularly effective for informal, short-form text such as Reddit posts.

Document embeddings were generated using the all-MiniLM-L12-v2 \cite{reimers2022} model, chosen for its balance of computational efficiency and semantic performance. Dimensionality reduction was performed using UMAP with cosine distance. Clustering was performed using HDBSCAN with a minimum cluster size of 70, reflecting the scale and density of the Reddit dataset.

Topic representations were initially extracted using a CountVectorizer with an n-gram range of (1, 2), and later refined to (3, 5) after topic reduction to capture more descriptive multi-word phrases. Maximal Marginal Relevance (MMR, diversity = 0.3) was applied to balance keyword relevance and diversity.

\subsection{Topic Representation and
Reduction}\label{topic-representation-and-reduction}

Initial topic discovery was performed with automatic topic selection. To improve interpretability, topics were reduced using BERTopic's \texttt{reduce\_topics} method, which merges semantically similar clusters. In this study, topics were reduced to 30, balancing granularity with interpretability.

Rather than merging large topic metadata directly into the main dataset, lightweight annotations were used. Each document was assigned:

\begin{itemize}
    \item a topic ID
    \item a topic probability score
    \item a binary flag indicating whether it is a representative document for its topic
\end{itemize}

This design avoids unnecessary data duplication while preserving interpretability and enabling efficient downstream analysis.

\subsection{Topic Evolution}\label{topic-evolution}

To analyse how discussions evolve over time, BERTopic's \texttt{topics\_over\_time} method was used with both \texttt{evolution\_tuning} and \texttt{global\_tuning} enabled to smooth temporal transitions.

Timestamps were derived from the \texttt{created\_utc} field and converted into datetime format. Documents were grouped into monthly time bins to reflect the long time span of the Reddit dataset. The number of bins was dynamically determined based on dataset duration (bounded between 10 and 50) to maintain consistent temporal resolution.

\subsection{Emotion Analysis}\label{emotion-analysis}

Emotion detection was performed using a pre-trained transformer model from the Hugging Face model hub. The pipeline assigns an emotion label to each document along with confidence scores.

The model was loaded using the \texttt{transformers} pipeline with a text-classification task configuration. GPU acceleration was used when available, with automatic fallback to CPU. Input texts were truncated to 512 tokens to comply with model limits.

Batch processing (batch size = 32) was used to improve performance. For each document, the model outputs scores across multiple emotion categories. The highest-confidence label was selected as the primary emotion, while all scores were retained for more detailed analysis.

If emotion labels were already present (from a prior classification step), they were reused to ensure consistency and avoid redundant computation.

Emotion analysis was performed using the RoBERTa-Base GoEmotions model \cite{lowe2022}, which predicts 28 emotion categories. For each document, the highest-confidence emotion label was retained as the primary label, while the three highest-scoring emotions were also stored for supplementary analysis.

\subsection{Visualisations}\label{visualisations}

\textbf{Word Clouds.}
Emotion-specific word clouds were generated using the primary emotion labels produced by the RoBERTa-base GoEmotions classifier. Documents sharing the same predicted emotion were grouped, and their text was aggregated into frequency-based visualisations to highlight the most characteristic vocabulary associated with each emotional category.

Each emotion was assigned a distinct colour map to improve interpretability, and word clouds were exported as high-resolution PNG images (1200 by 600 pixels) containing up to 100 words.

\textbf{Emotion Time-Series.} Interactive time-series visualisations were generated using the predicted GoEmotions labels. Emotion frequencies were aggregated over monthly intervals for the Reddit dataset and displayed using two complementary views:

\begin{itemize}
    \item Percentage view, showing the relative proportion of each emotion over time.
    \item Stacked count view, showing the absolute number of posts assigned to each emotion over time.
\end{itemize}

The visualisations were implemented using Plotly to support interactive exploration of temporal emotional trends.

\textbf{Topic Evolution.} Dynamic topic trends were visualised using BERTopic's \texttt{visualize\_topics\_over\_time} function. These interactive visualisations display topic prevalence as time series, allowing identification of emerging, stable, and declining themes throughout the study period.

\subsection{Keyword and Community Analysis}\label{keyword-and-community-analysis}

\textbf{Topic-Based Community Structure.} Communities are defined using the outputs of BERTopic clustering. Each document is assigned both a topic and a reduced (core) topic, along with associated probability scores (\texttt{topic\_proba}, \texttt{core\_topic\_proba}) representing model confidence.

Topic reduction is performed using \texttt{topic\_model.$reduce\_topics$(...,$nr\_topics=nr\_topics$)} to combine semantically similar topics into a smaller set (30), forming higher-level thematic communities. Each core topic therefore represents a cluster of related discussions, enabling structured exploration of large-scale discourse.

\textbf{Representative Documents.} To preserve interpretability without increasing dataset size, representative documents are identified using a flagging approach. A document is marked as representative if its \texttt{cleaned\_text} appears in the topic’s \texttt{Representative\_Docs} list.

These documents act as anchors for each community, capturing its most characteristic language while avoiding storage of large text fields. This enables efficient analysis while retaining interpretability.

\textbf{Keyword Extraction.} Keywords are derived implicitly from topic representations rather than stored as standalone fields. After topic reduction, topic descriptors were updated using \texttt{topic\_model\_clustered.update\_topics(..., n\_gram\_range=(3, 5))} to produce multi-word expressions (3–5 grams) that served as topic descriptors. Compared to single words, these phrases capture more context, reduce ambiguity, and better reflect natural language usage.

These keywords are embedded in topic metadata and are used for labelling, visualisation, and interpretation of community themes.

\textbf{Emotion-Enriched Keywords.} Emotion labels provide an additional layer of analysis by linking language patterns to affective states. Each document is assigned one of the 28 emotion categories predicted by the RoBERTa-base GoEmotions classifier \cite{lowe2022}. These labels provide an additional layer of semantic interpretation by linking language patterns with fine-grained emotional states.

For each emotion, documents sharing the same predicted label were aggregated to generate emotion-specific word clouds. These visualisations highlight the vocabulary most strongly associated with each emotional category, enabling qualitative comparison of language across different emotional contexts.

For each emotion, subsets of documents are used to generate word clouds. These act as emotion-conditioned keyword distributions, highlighting the most salient terms associated with each emotional category. This enables comparison of how different themes are expressed across emotional contexts.

\textbf{Temporal Dynamics.} Community dynamics are examined through temporal emotion trends. Documents are aggregated into monthly intervals based on \texttt{created\_utc}, and emotion distributions are computed as:

\begin{itemize}
    \item Percentage of posts over time
    \item Absolute counts (stacked)
\end{itemize}

These trends provide a proxy for how thematic communities evolve, revealing shifts in sentiment, periods of heightened emotional response, and longer-term changes in discourse.

\textbf{Summary.} Communities are defined as clusters of semantically related documents derived from topic modelling, while keywords are represented through n-gram topic descriptors and emotion-conditioned term distributions. Together, these components enable structured exploration of large text corpora and provide insight into how climate-related discussions are organised, expressed, and evolve over time.

\section{Results \& Discussion}\label{results and discussion}

Although the overall ClimateLens framework is designed for both Reddit and Twitter/X, this preprint focuses on Reddit for a more in-depth analysis that fits the current project timeline.

\subsection{Topic modelling Results}\label{topic-modelling-results}

Topic modelling was used to identify the main themes present in climate-related discussions. Using BERTopic, we extracted a set of distinct topics and analysed how they differ in content, structure, and prominence across Reddit communities. The model identified a range of topics that can be broadly grouped into a few recurring themes:

\begin{itemize}
    \item Action-oriented discussions, such as climate advocacy, policy, and collective mobilisation
    \item Informational content, focused on explaining climate change and its causes
    \item Personal and emotional responses, where users express concern, frustration, or uncertainty
\end{itemize}

The presence of these themes suggests that climate-related discourse is not solely focused on environmental issues but also reflects broader social and psychological responses to climate change. Discussions frequently move beyond scientific facts and into questions of responsibility, action, and emotional coping. This indicates that climate anxiety is intertwined with perceptions of agency, community engagement, and trust in institutions.

\textbf{Topic Structure and
Relationships.} Analysis of topic relationships shows that discussions are not evenly distributed, but instead cluster into a small number of broader thematic groups.

The topic structure suggests two broad clusters:

\begin{itemize}
    \item One cluster is centred around action and engagement, including advocacy efforts, political discourse, and organised responses to climate issues.
    \item The second cluster focuses on information and reflection, including general awareness, explanations, and personal interpretations of climate-related topics.
\end{itemize}

Topics within each cluster share similar vocabulary and framing, while topics across clusters are more distinct. One interpretation is that conversations about climate change naturally separate into two distinct modes of engagement: discussing and understanding climate issues versus organising around action and advocacy. While these groups share an overarching concern about climate change, they appear to use different language and emphasise different objectives.

The separation between these clusters has important implications. Although increasing public awareness is essential for climate action, our findings suggest that awareness-oriented and action-oriented discussions often remain distinct within online communities. This disconnect points to a gap between awareness and engagement, highlighting the need for communication strategies that more effectively link climate education with opportunities for action.

\textbf{Dynamic Topic modelling.} Dynamic topic modelling was used to examine the weight of different climate-related discussions over time. Rather than remaining static, online conversations fluctuate in response to external events, evolving public interest, and community engagement.

\begin{figure}[h] 
    \centering
    \includegraphics[width=1.0\textwidth]{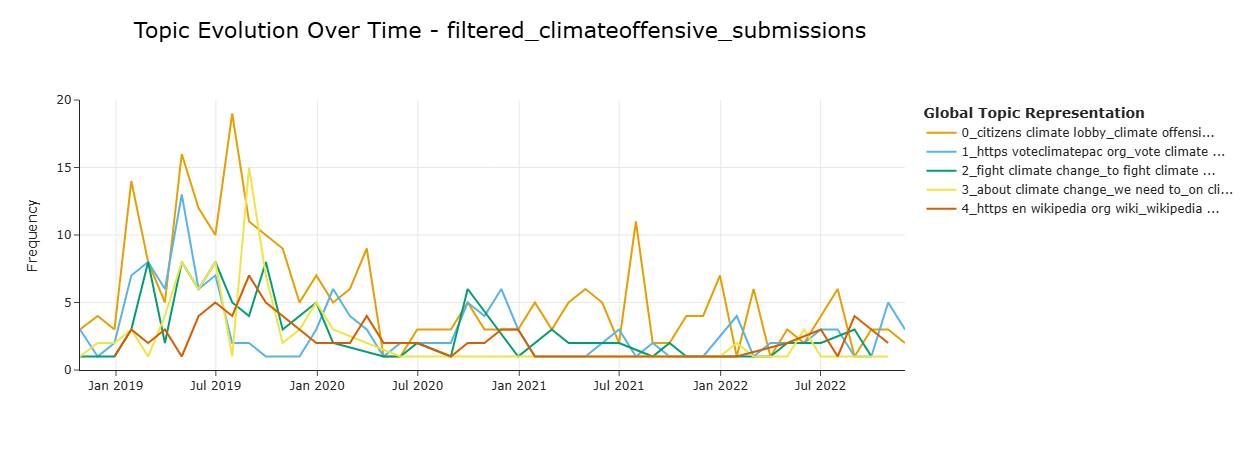}
    \caption{Representative topic evolution for submissions from r/climateoffensive's}
    \label{fig:topic-dtm}
\end{figure}

The temporal variation observed across topics highlights the dynamic nature of online climate discourse. Periods of heightened activity may correspond to moments when public attention toward climate issues is elevated, creating opportunities for policymakers, educators, and advocacy organisations to reach larger audiences. Conversely, sustained declines in activity illustrate the challenge of maintaining long-term engagement with climate-related issues after major events fade from public attention.

To illustrate how these discussions evolved over time, we examine the keywords associated with two of the five topics across the study period. For brevity, we present only two topics here, while the interactive visualization in Appendix~\ref{appendix} includes all five shown above. This temporal perspective provides additional context for interpreting the peaks and declines shown in Figure~\ref{fig:topic-dtm}.

Topic 0 (orange) captures a broad range of discussions related to climate policy, advocacy, finance, and environmental organisations. Representative keywords include phrases such as "to help climate," "our work urgent projects," and "bill to combat climate," alongside references to the White House, and financial discussions involving central banks and economic resources. The topic also frequently references organisations including Citizens' Climate Lobby, GiveMN, The Rainforest Trust, and The Intrepid Foundation, with mentions of Citizens' Climate Lobby and The Intrepid Foundation increasing notably around July 2019, corresponding to the highest peak in topic prevalence.

Topic 2 (green) is more strongly associated with environmental activism and conservation initiatives. Early peaks coincide with frequent references to organisations such as The Intrepid Foundation and the Global Meadows Movement, while a later increase around October 2020 highlights terms including "Ecosia 80," "biodiversity hotspots," and "reforestation projects." These keywords suggest increased discussion surrounding large-scale conservation and reforestation efforts, including Ecosia's commitment to directing a substantial portion of its profits toward global tree-planting initiatives.

\textbf{Intertopic Distance Map.} The intertopic distance map provides a high level view of the semantic relationships among discovered topics. Rather than forming a single continuous cluster, posts on r/climateoffensive separate into two primary groups. One group is centred on advocacy, collective action, and political engagement, and another is focused on information, education, and personal reflection. This division indicates that climate conversations frequently occur within distinct communicative contexts instead of blending into a single unified discussion space.

\begin{figure}[H] 
    \centering
    \includegraphics[width=0.8\textwidth]{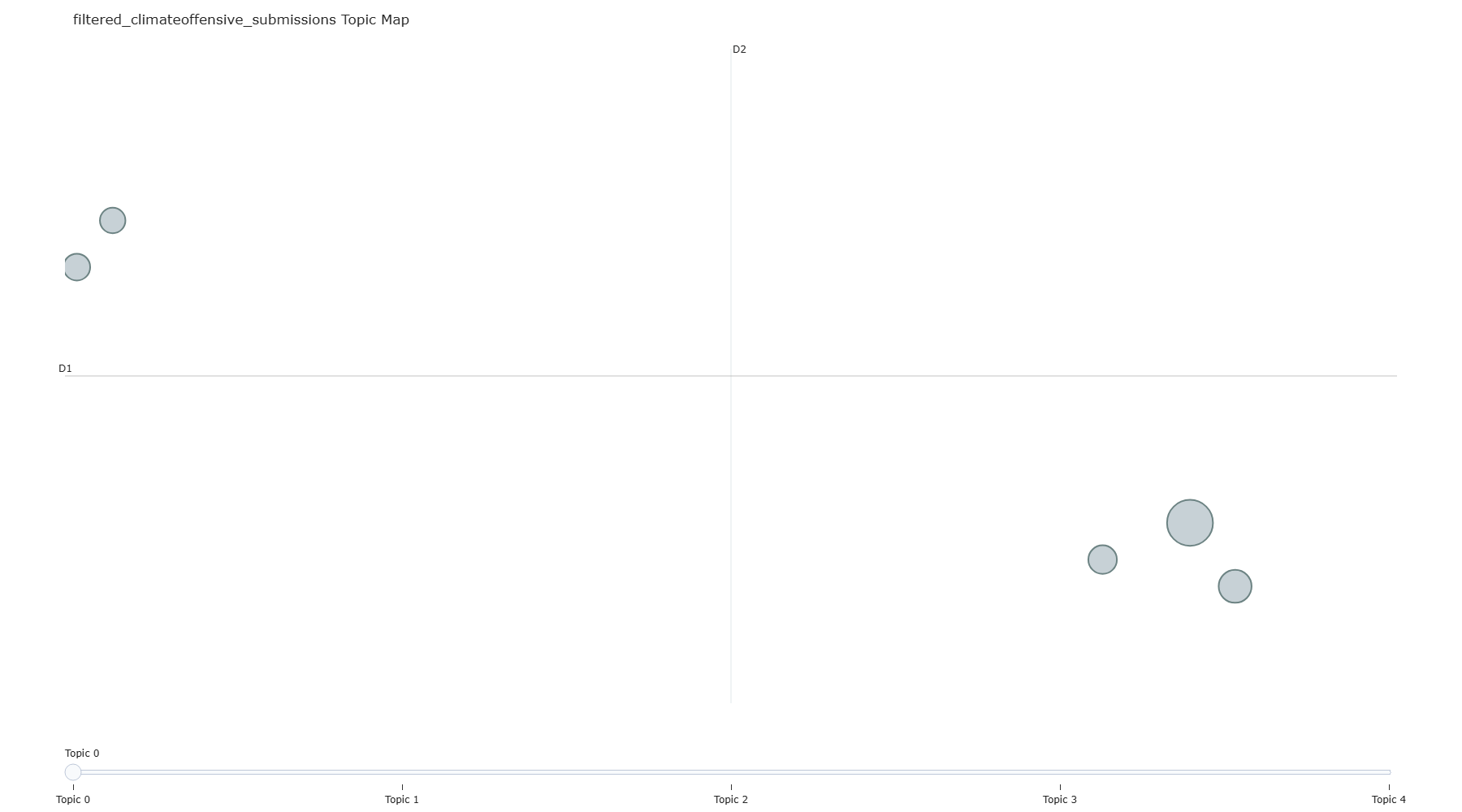}
    \caption{Representative Intertopic Distance Map of r/climateoffensive}
    \label{fig:topic-distance}
\end{figure}

\textbf{Topic Frequency Analysis} Keyword frequency analysis highlights the language that distinguishes different climate-related discussions. Examining the highest-weighted terms within each topic provides insight into what users emphasise and how conversations evolve between original posts and subsequent community responses.

Climate-related posts frequently focus on understanding climate change, its causes, and its consequences. In contrast, comments often contain stronger emotional language, calls for action, and expressions of frustration.

This difference suggests that while users initially engage with climate discussions to seek or share information, subsequent interactions often become emotionally driven. Such emotional engagement may help mobilise participation and awareness but may also contribute to polarisation if frustration and anger dominate discussions.

Overall, these findings suggest that effective climate communication should balance information sharing with constructive opportunities for engagement, helping users move from awareness toward action without amplifying emotional distress.

\textbf{Gettingoverit.} Conversations within r/gettingoverit present a notably different interaction pattern. Although many original posts describe emotional struggles, comments consistently provide encouragement, validation, and reassurance. Compared with other communities examined, this subreddit demonstrates stronger evidence of constructive and supportive engagement.

\begin{figure}[H] 
    \centering
    \includegraphics[width=0.7\textwidth]{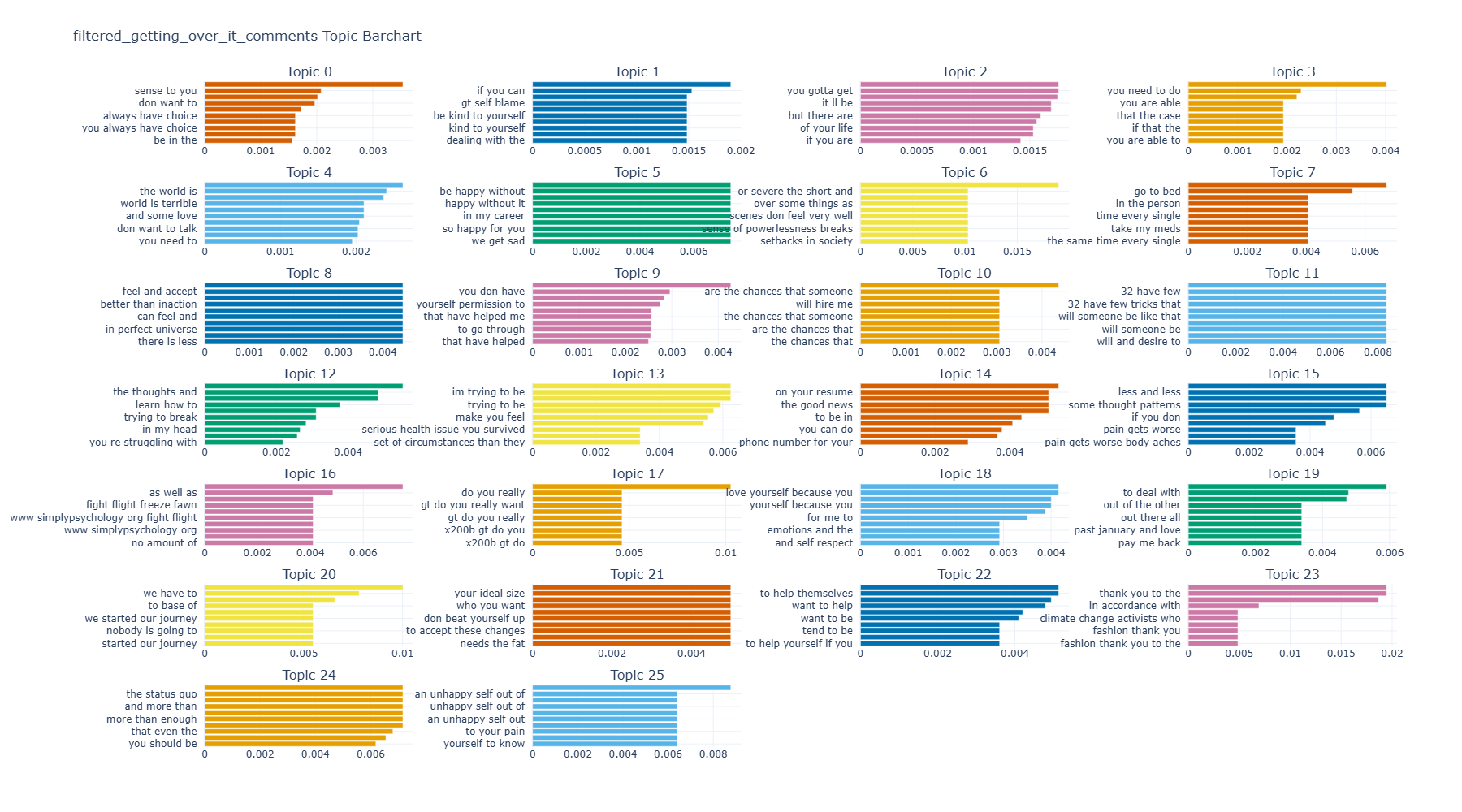}
    \caption{Representative barchart of topics discussed in comments on r/gettingoverit}
    \label{fig:barchart-gettingoverit}
\end{figure}

\textbf{Depression.} Discussions related to depression exhibit substantially higher levels of emotional distress than climate-focused discussions. Posts frequently express uncertainty and hopelessness, whereas comments vary between supportive responses and broader discussions that do not always directly address the author's concerns. This contrast highlights the challenges of providing meaningful peer support in online mental health communities.

\textbf{climateoffensive.} Discussions within r/climateoffensive primarily emphasise understanding climate change and encouraging collective action. While original posts often focus on climate issues themselves, comments more frequently introduce urgency, advocacy, and frustration regarding perceived inaction. Together, these patterns illustrate how informational discussions often transition into action-oriented community engagement.

\subsection{Emotion Patterns}\label{emotion-and-sentiment-patterns}

Unlike the topic modelling results, which present representative subreddit-level analyses, emotion classification was performed across the complete Reddit corpus to identify broader emotional patterns. Representative emotion distributions for selected communities are shown below, while corpus-level vocabulary summaries are presented in the subsequent word cloud analysis.

Emotion analysis was conducted to examine how individuals express and respond to climate-related discussions on Reddit. Each post and comment was assigned an emotion label using the RoBERTa-base GoEmotions model \cite{lowe2022}, enabling analysis of emotional distributions across topics and communities.

Overall, emotional expression across the dataset is not uniform. Certain topics are associated more strongly with negative emotions such as fear, sadness, disappointment, and grief, while others exhibit more balanced emotional profiles. In general, original posts tend to express stronger negative emotional signals, whereas comments often introduce a wider range of responses, including support, reassurance, curiosity, and neutral reactions.

Collectively, these findings indicate that climate anxiety is expressed through a diverse range of emotional experiences rather than being limited to fear or sadness alone. Community responses further suggest that emotional expression evolves through discussion and interaction, highlighting the complexity of emotional engagement within climate-related conversations.

\textbf{Topic-Level Emotional
Patterns.} Emotional distributions vary substantially across discussion topics, illustrating that different online communities express and respond to climate-related concerns in distinct ways. To illustrate these differences, we present representative examples drawn from several Reddit communities that capture a range of emotional profiles. 

Climate-focused discussions on r/climateoffensive are more likely to show neutral emotional expression overall, with fewer strong emotional signals compared to mental health–related topics. This may reflect a more informational or discussion-based tone, though occasional spikes in frustration or concern are still present.

\begin{figure}[H]
    \centering
    \begin{minipage}{0.45\textwidth}
        \centering
        \includegraphics[width=0.9\textwidth]{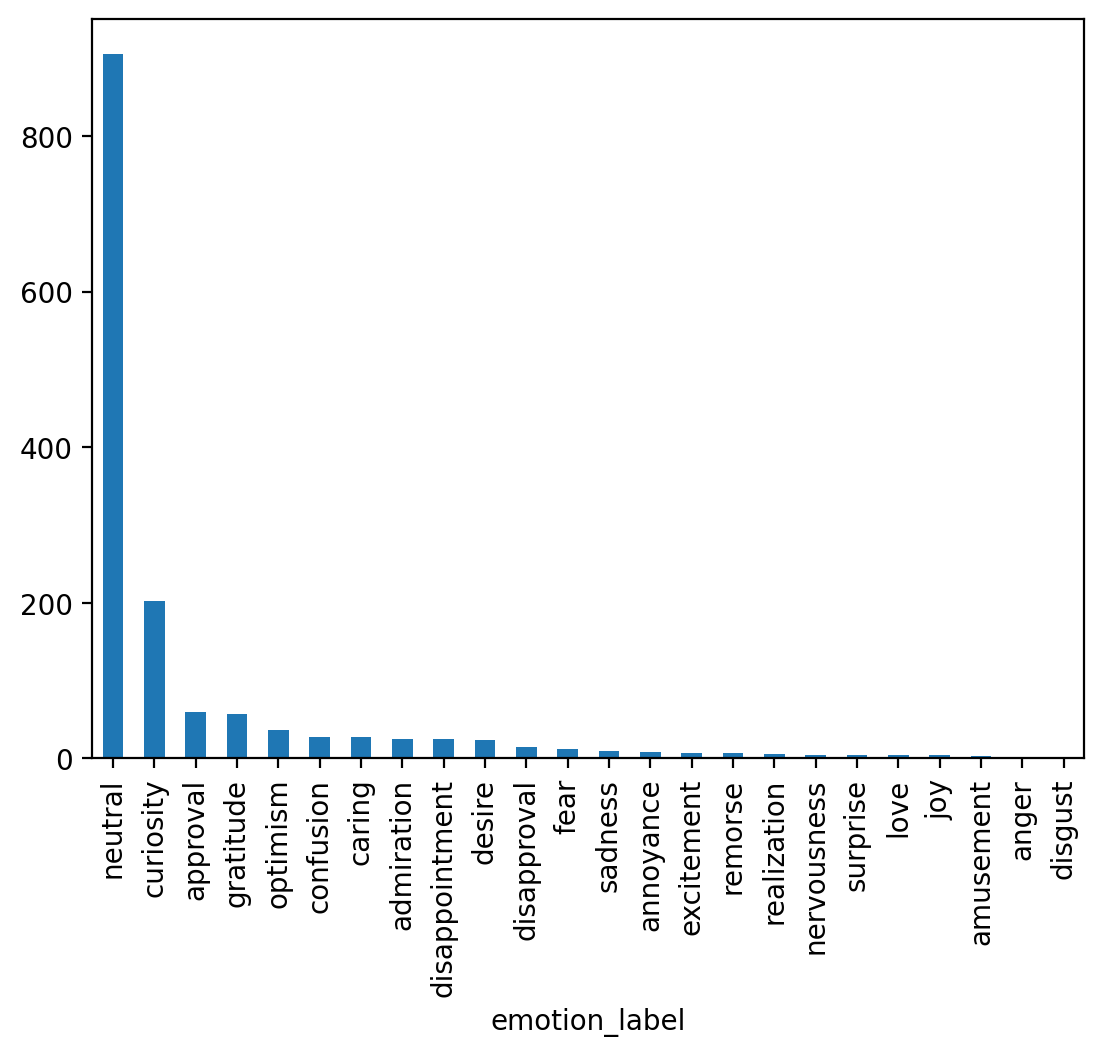} 
        \caption{Emotion distribution in submissions from r/climateoffensive}
    \end{minipage}\hfill
    \begin{minipage}{0.45\textwidth}
        \centering
        \includegraphics[width=0.9\textwidth]{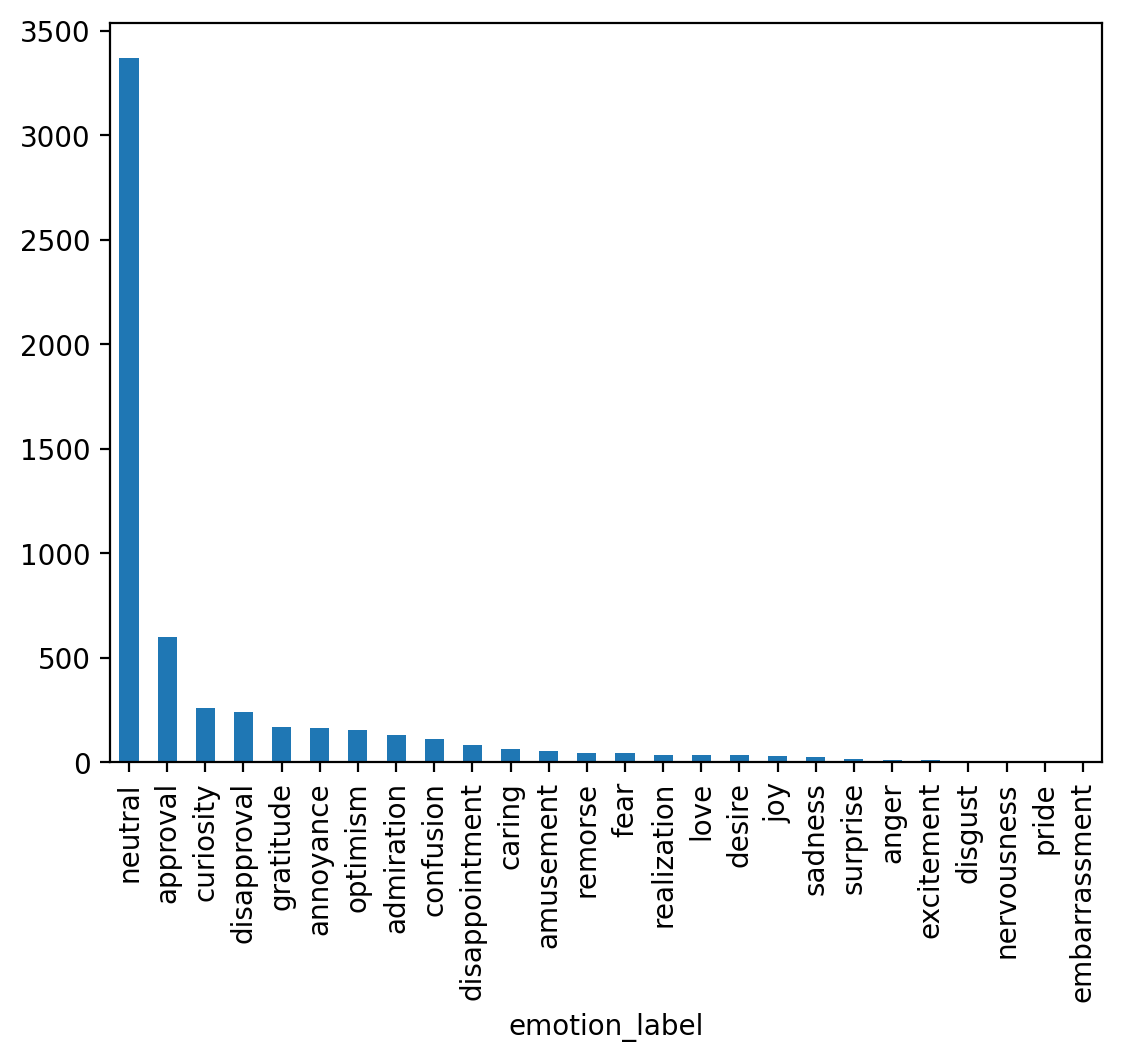} 
        \caption{Emotion distribution in comments from r/climateoffensive}
    \end{minipage}
\end{figure}

Anxiety-related discussions are primarily characterised by fear, nervousness, and sadness. These posts often reflect uncertainty and concern about the future. In contrast, comments on these posts include more caring and neutral responses, suggesting that other users often respond with empathy rather than amplifying distress.

\begin{figure}[!htbp]
    \centering
    \begin{minipage}{0.45\textwidth}
        \centering
        \includegraphics[width=0.9\textwidth]{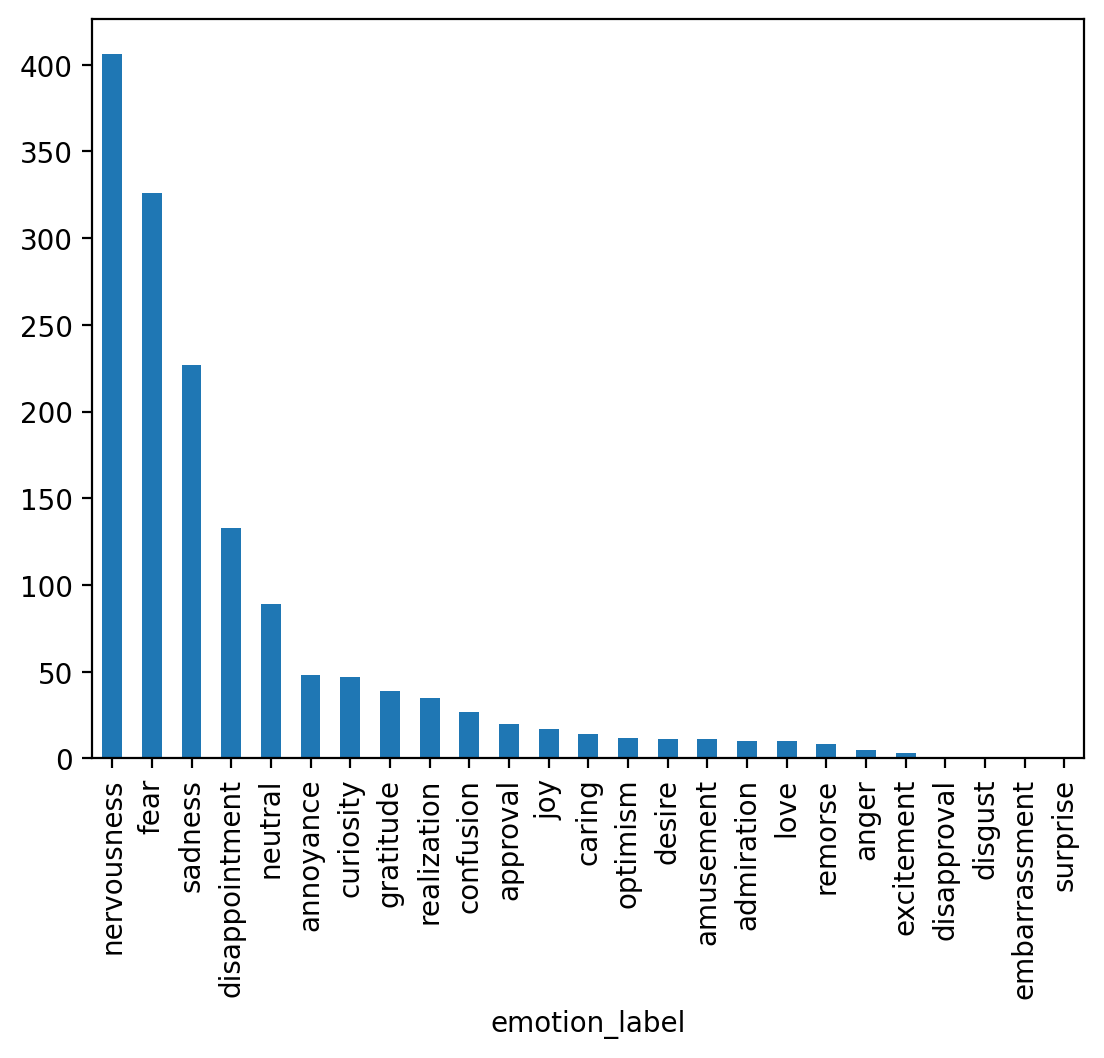} 
        \caption{Emotion distribution in submissions from r/anxiety.}
    \end{minipage}\hfill
    \begin{minipage}{0.45\textwidth}
        \centering
        \includegraphics[width=0.9\textwidth]{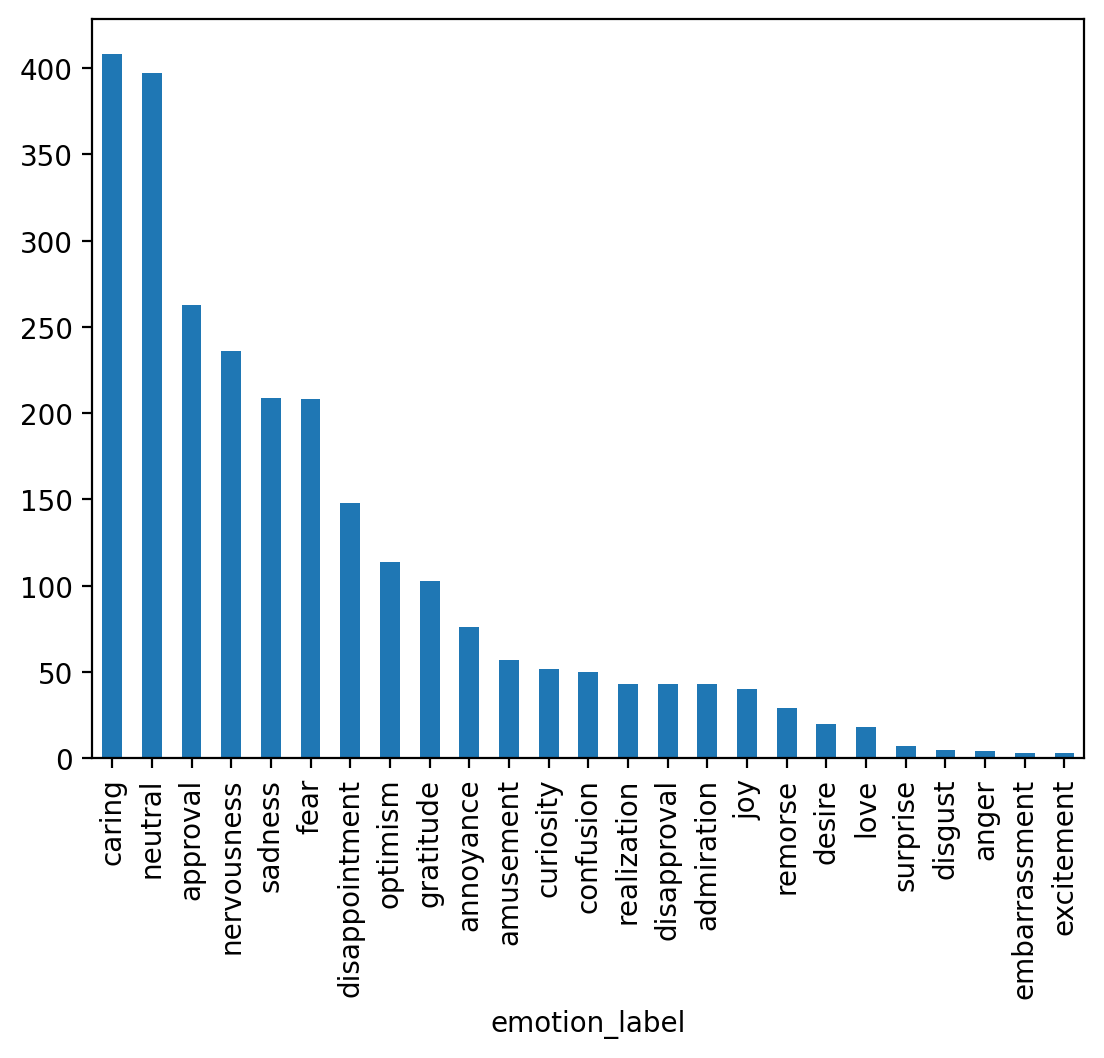} 
        \caption{Emotion distribution in comments from r/anxiety.}
    \end{minipage}
\end{figure}

Depression-related topics show a strong concentration of sadness and disappointment in both posts and comments. However, comments also include a mix of supportive emotions such as care and occasional optimism, indicating attempts by the community to provide reassurance even when the original content is highly negative.

\begin{figure}[H]
    \centering
    \begin{minipage}{0.45\textwidth}
        \centering
        \includegraphics[width=0.9\textwidth]{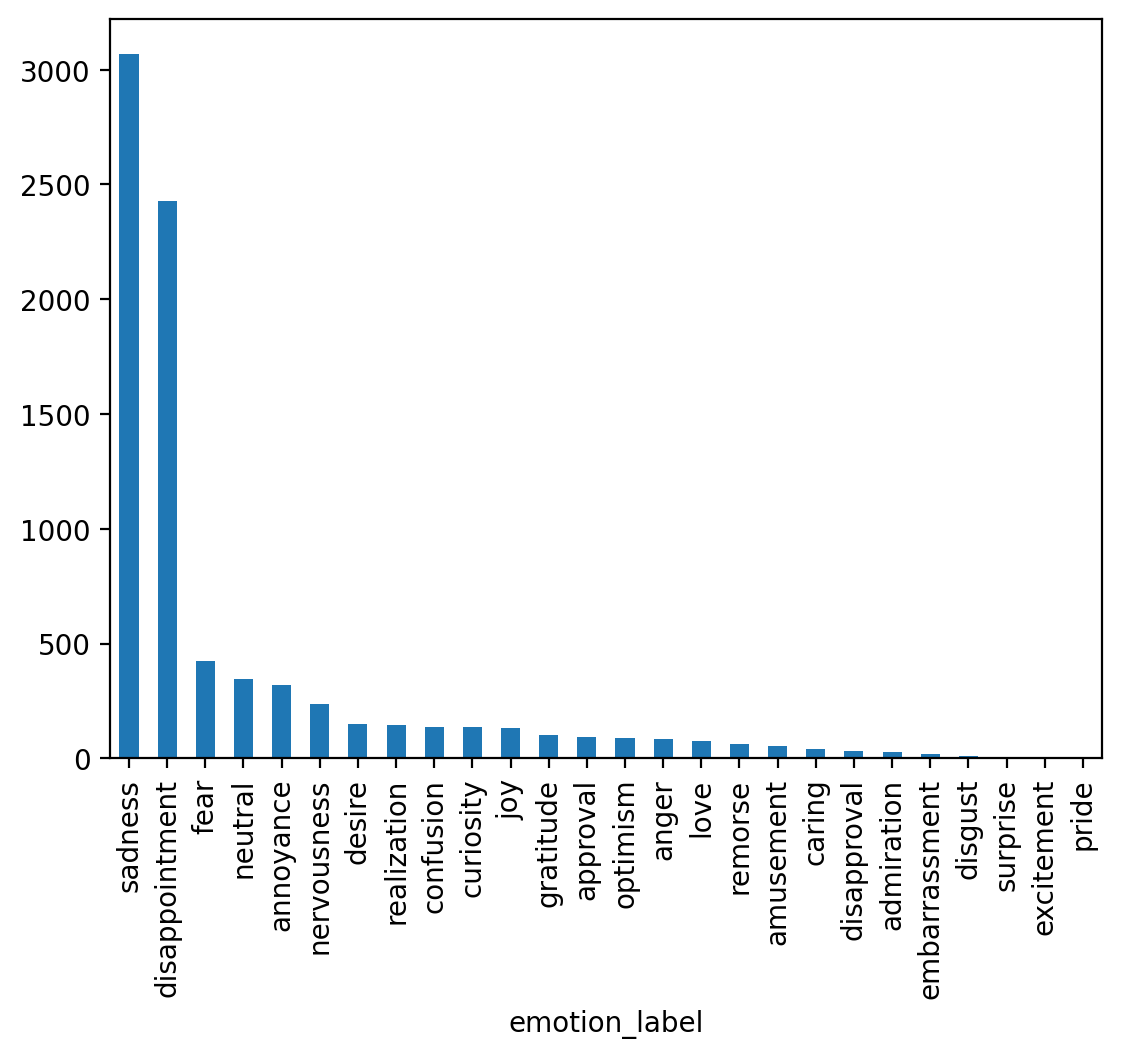} 
        \caption{Emotion distribution in submissions from r/depression}
    \end{minipage}\hfill
    \begin{minipage}{0.45\textwidth}
        \centering
        \includegraphics[width=0.9\textwidth]{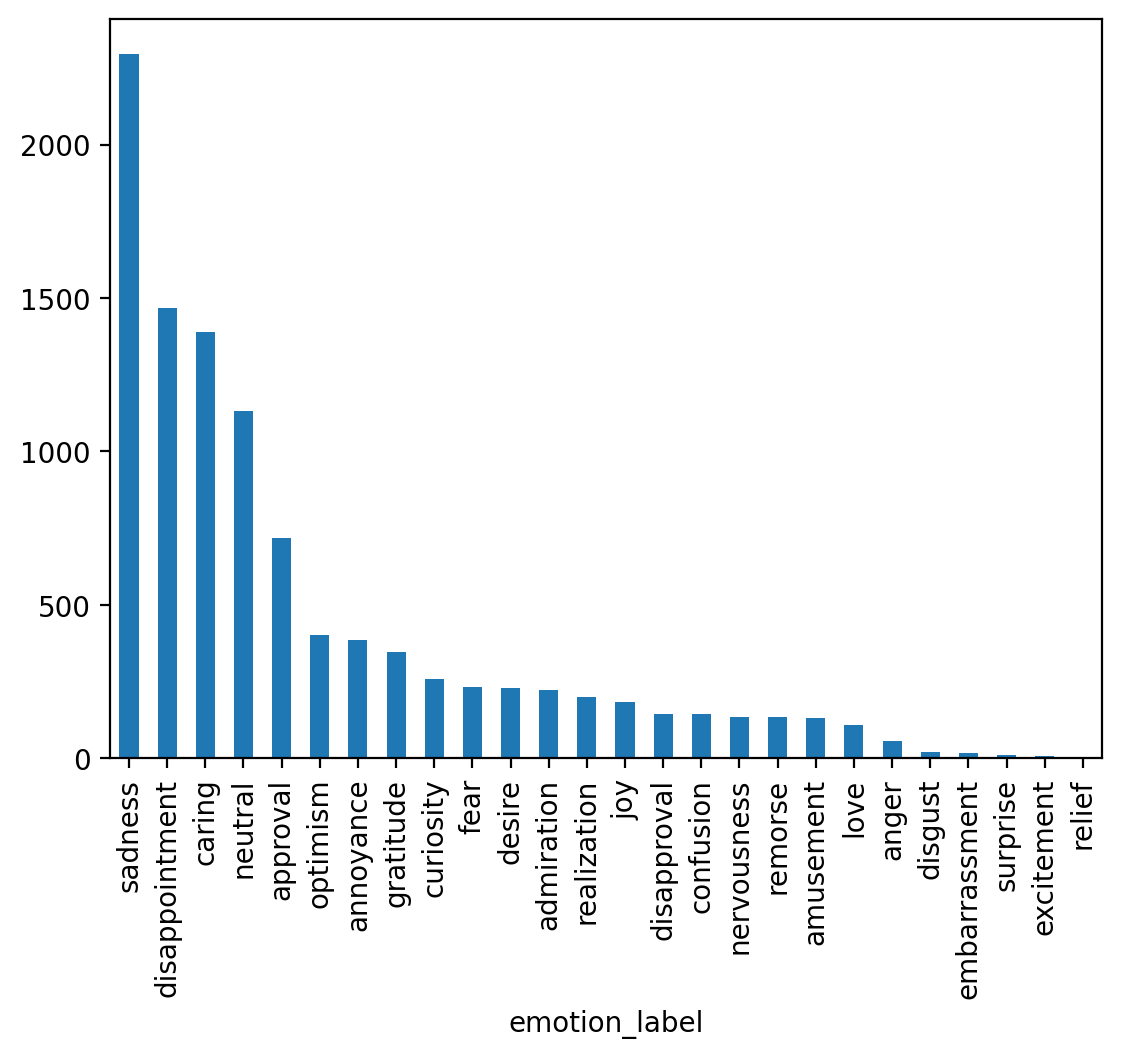} 
        \caption{Emotion distribution in comments from r/depression}
    \end{minipage}
\end{figure}

Discussions on r/gettingoverit display a more balanced emotional distribution. Posts frequently contain sadness and hesitation, while comments are more consistently supportive, with expressions of encouragement, validation, and gratitude. This suggests a more constructive interaction pattern compared to other topics.

\begin{figure}[H]
    \centering
    \begin{minipage}{0.45\textwidth}
        \centering
        \includegraphics[width=0.9\textwidth]{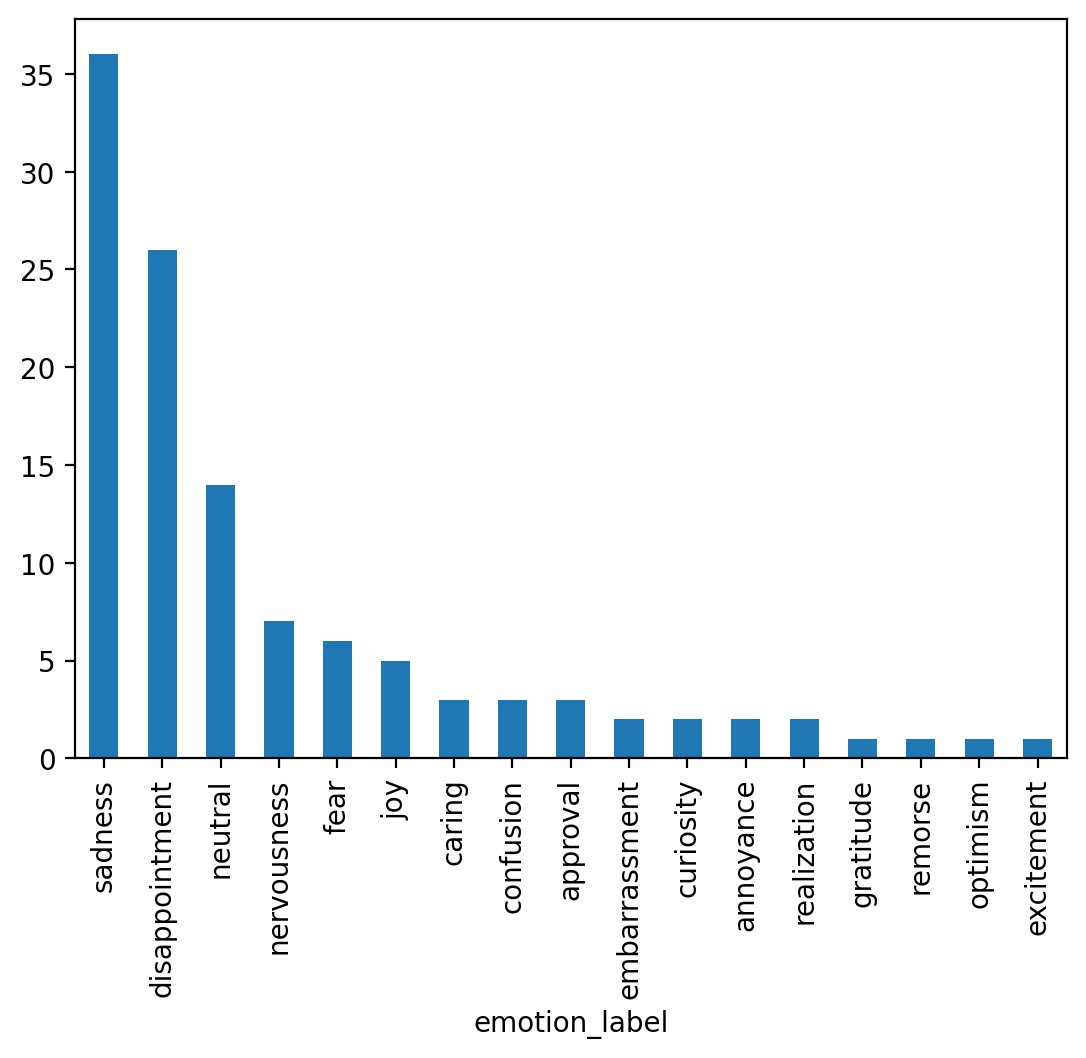} 
        \caption{Emotion distribution in submissions from r/gettingoverit}
    \end{minipage}\hfill
    \begin{minipage}{0.45\textwidth}
        \centering
        \includegraphics[width=0.9\textwidth]{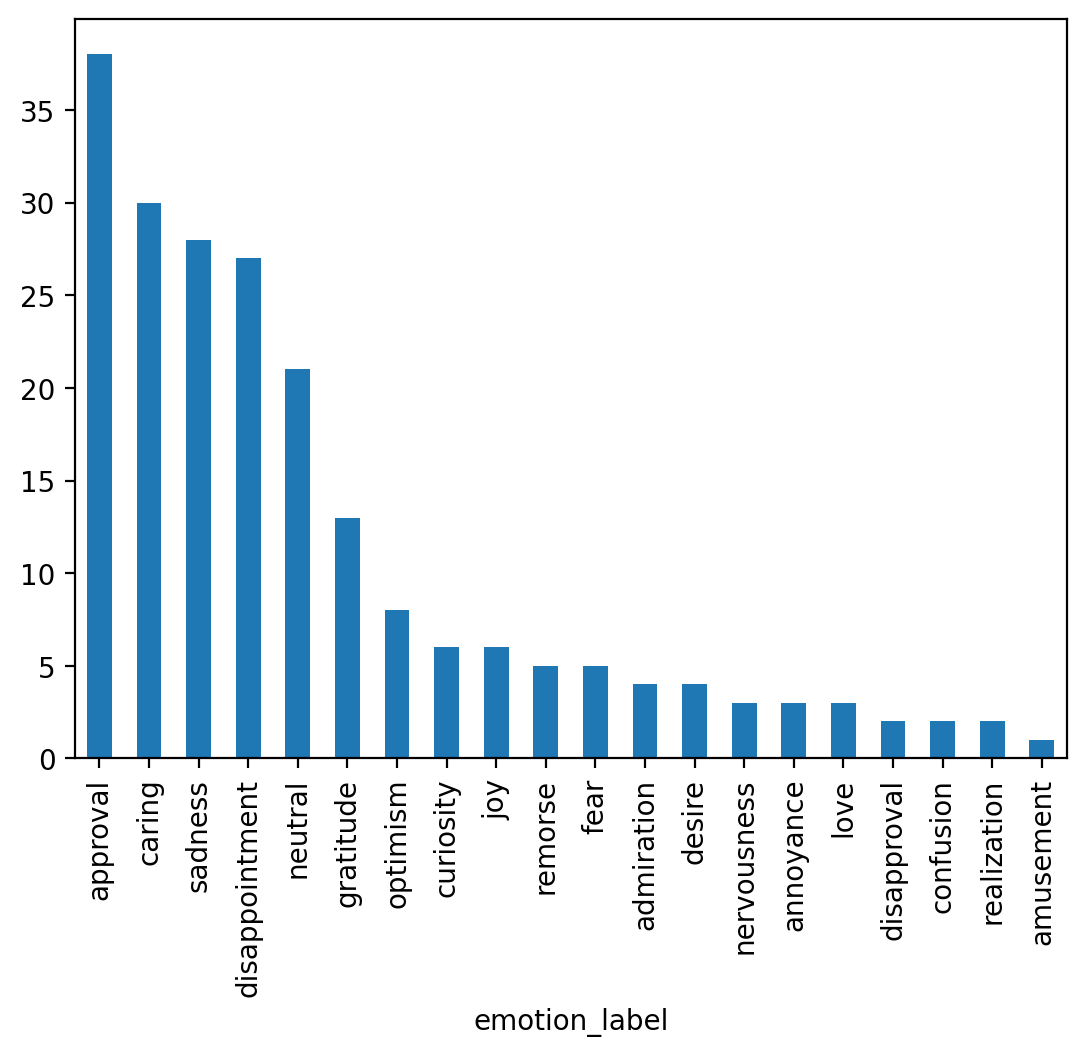} 
        \caption{Emotion distribution in comments from r/gettingoverit}
    \end{minipage}
\end{figure}

Teen-related discussions on r/teenagers are largely neutral, but include smaller proportions of annoyance, curiosity, and sadness. This indicates that while most interactions are not strongly emotional, certain conversations still carry underlying emotional depth.

\begin{figure}[H]
    \centering
    \begin{minipage}{0.45\textwidth}
        \centering
        \includegraphics[width=0.9\textwidth]{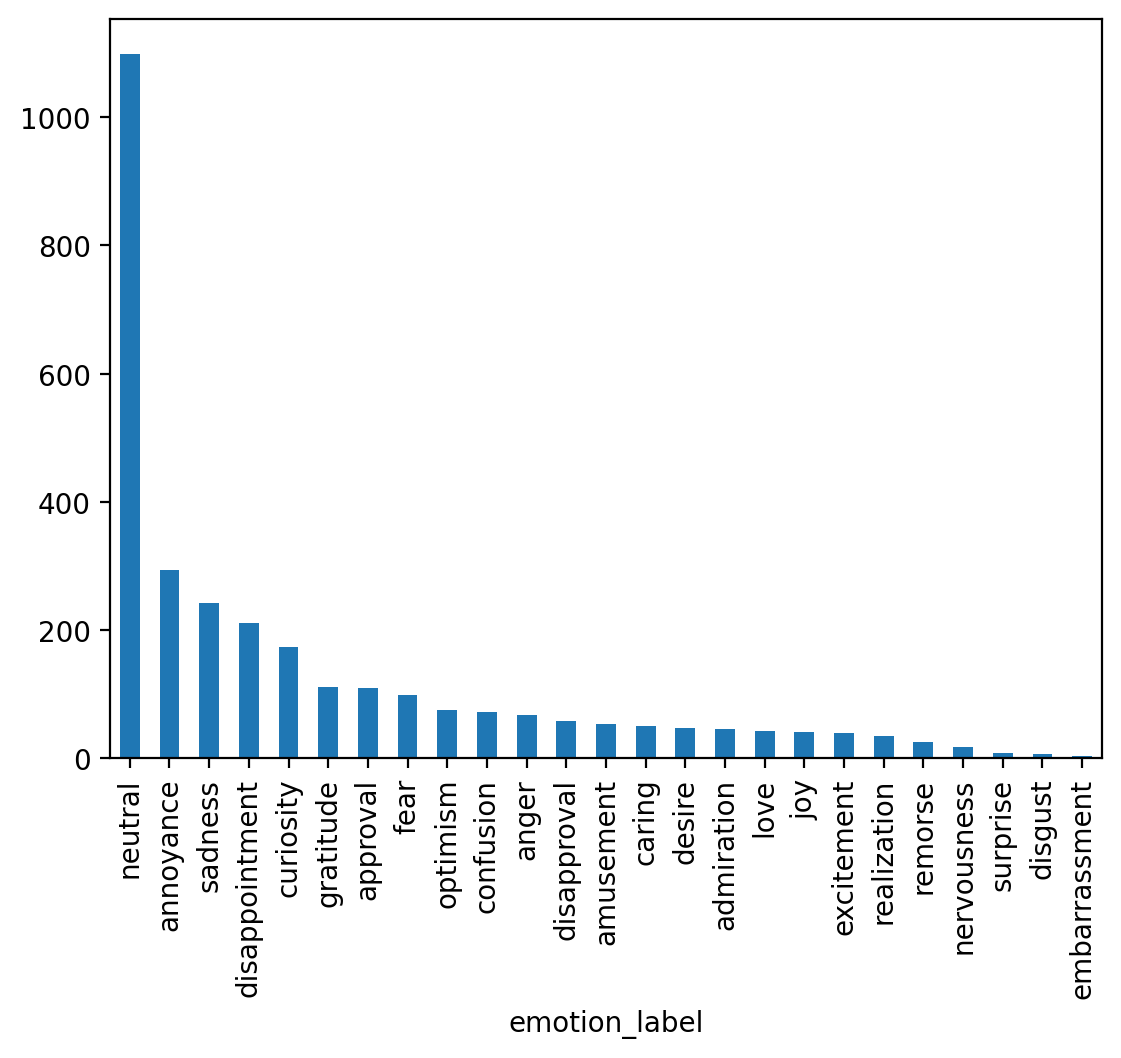} 
        \caption{Emotion distribution in submissions from r/teenagers}
    \end{minipage}\hfill
    \begin{minipage}{0.45\textwidth}
        \centering
        \includegraphics[width=0.9\textwidth]{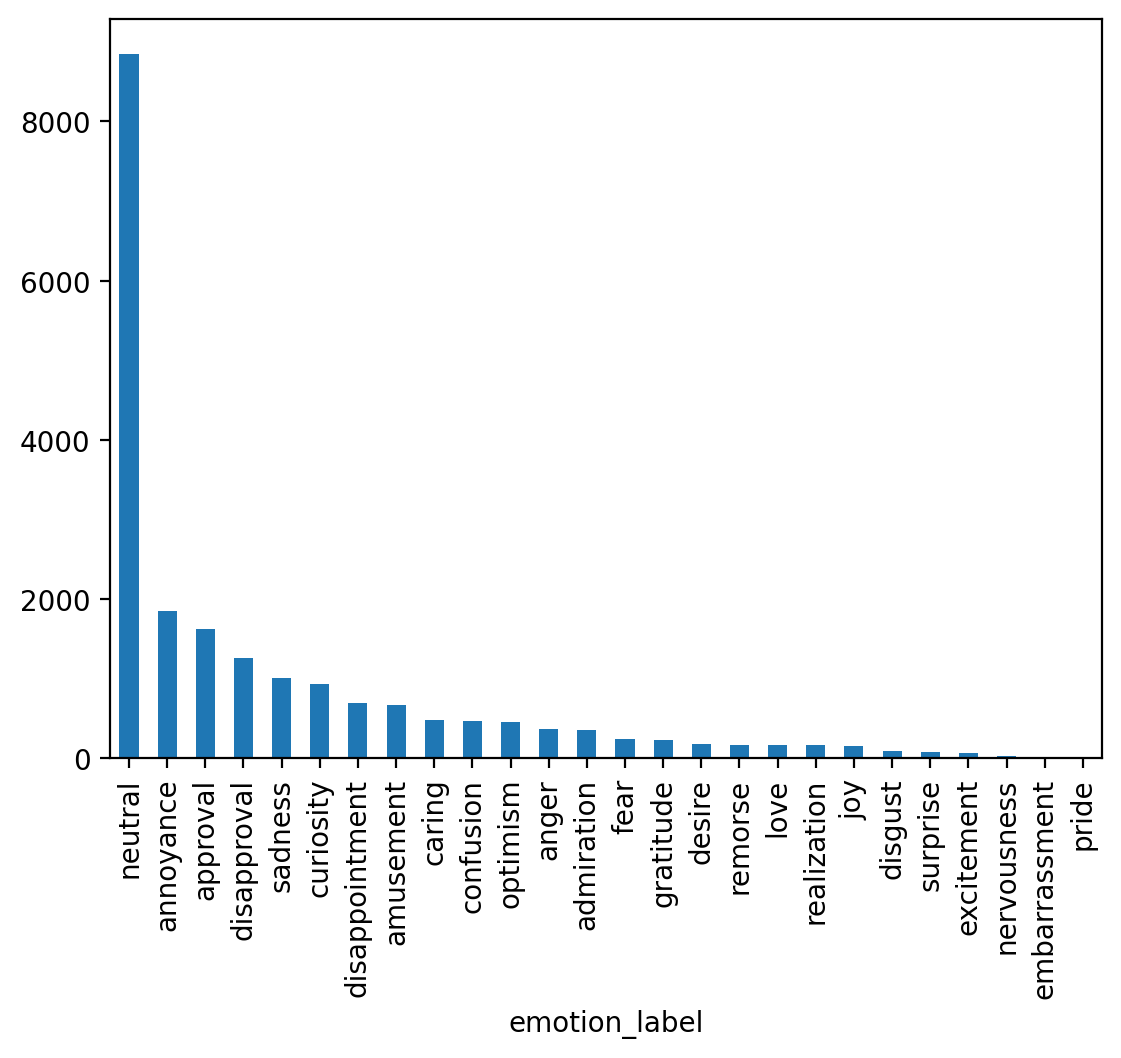} 
        \caption{Emotion distribution in comments from r/teenagers}
    \end{minipage}
\end{figure}

\textbf{Post vs. Comment Dynamics.} A consistent pattern across multiple topics is the difference between posts and comments. Posts are more likely to express strong negative emotions while comments tend to introduce a broader range of responses.

In several cases, comments shift the tone of the discussion by adding:

\begin{itemize}
    \item Care and empathy (reassurance, validation)
    \item Neutral or stabilising responses
    \item Occasional positive signals such as encouragement or gratitude
\end{itemize}

These patterns indicate that Reddit discussions frequently function as interactive emotional exchanges rather than isolated expressions of distress. Community members often introduce empathy, reassurance, or alternative perspectives that reshape the emotional tone of the conversation.

\subsection{Word-Level Emotional
Signals}\label{word-level-emotional-signals}

\begin{figure}[H]
\centering

\begin{subfigure}{0.65\textwidth}
    \centering
    \includegraphics[width=\linewidth]{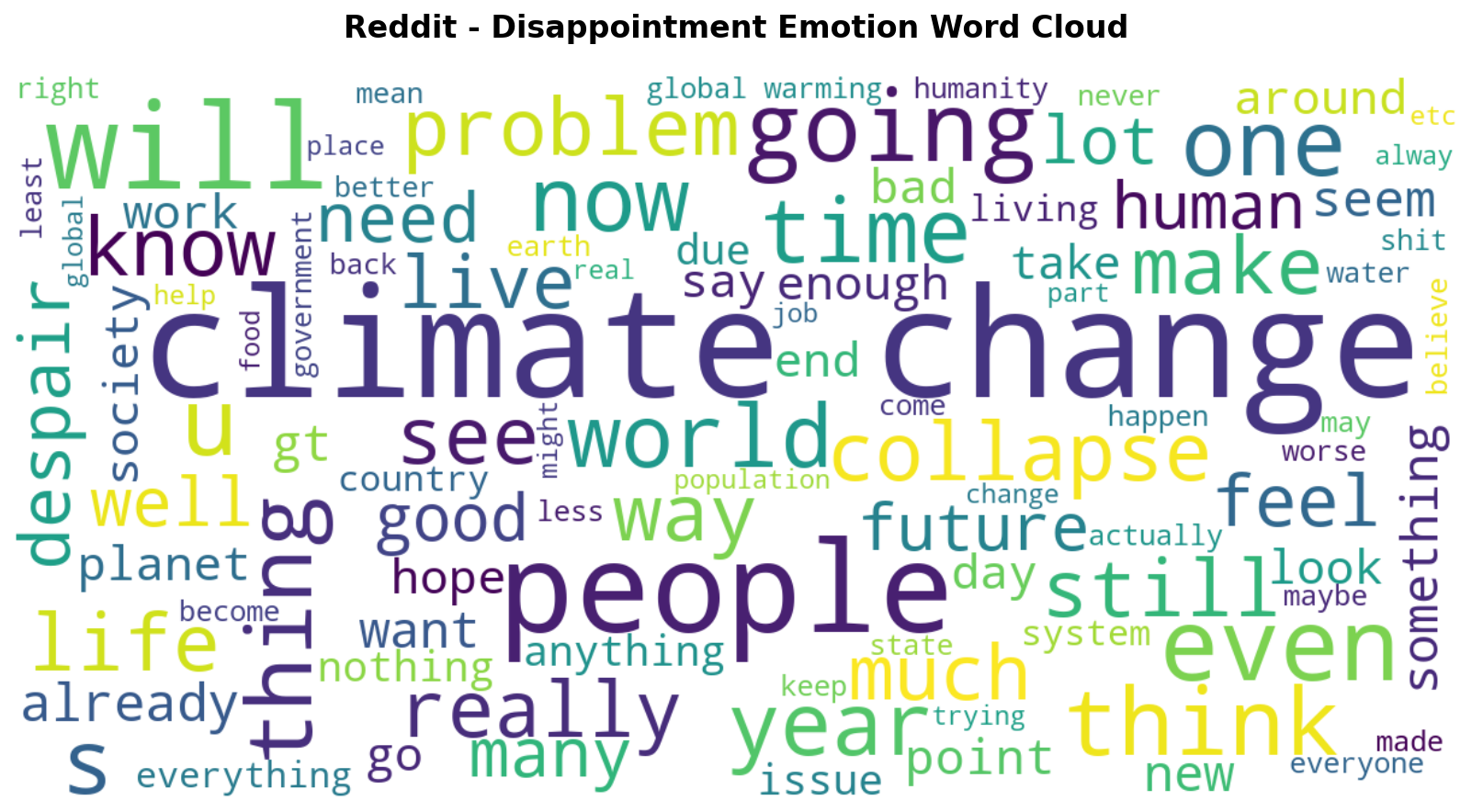}
    \caption{Disappointment}
\end{subfigure}

\caption{Word cloud illustrating representative vocabulary associated with disappointment posts.}
\label{fig:wordclouds-page2}
\end{figure}

\textbf{Content Warning.} The following figures reproduce words extracted directly from Reddit posts. Some word clouds contain profanity and language relating to depression, self-harm, emotional distress, and other sensitive topics. These terms are presented solely to illustrate patterns of emotional expression in the dataset and do not reflect the views of the authors. Readers may wish to exercise discretion, especially when viewing Figure~\ref{fig:wordclouds-page2}.

Word cloud visualisations were generated by aggregating documents across the complete Reddit dataset according to their predicted emotion labels. They highlight commonly associated vocabulary within each emotion category.

\begin{figure}[H]
\centering

\begin{subfigure}{0.48\textwidth}
    \centering
    \includegraphics[width=\linewidth]{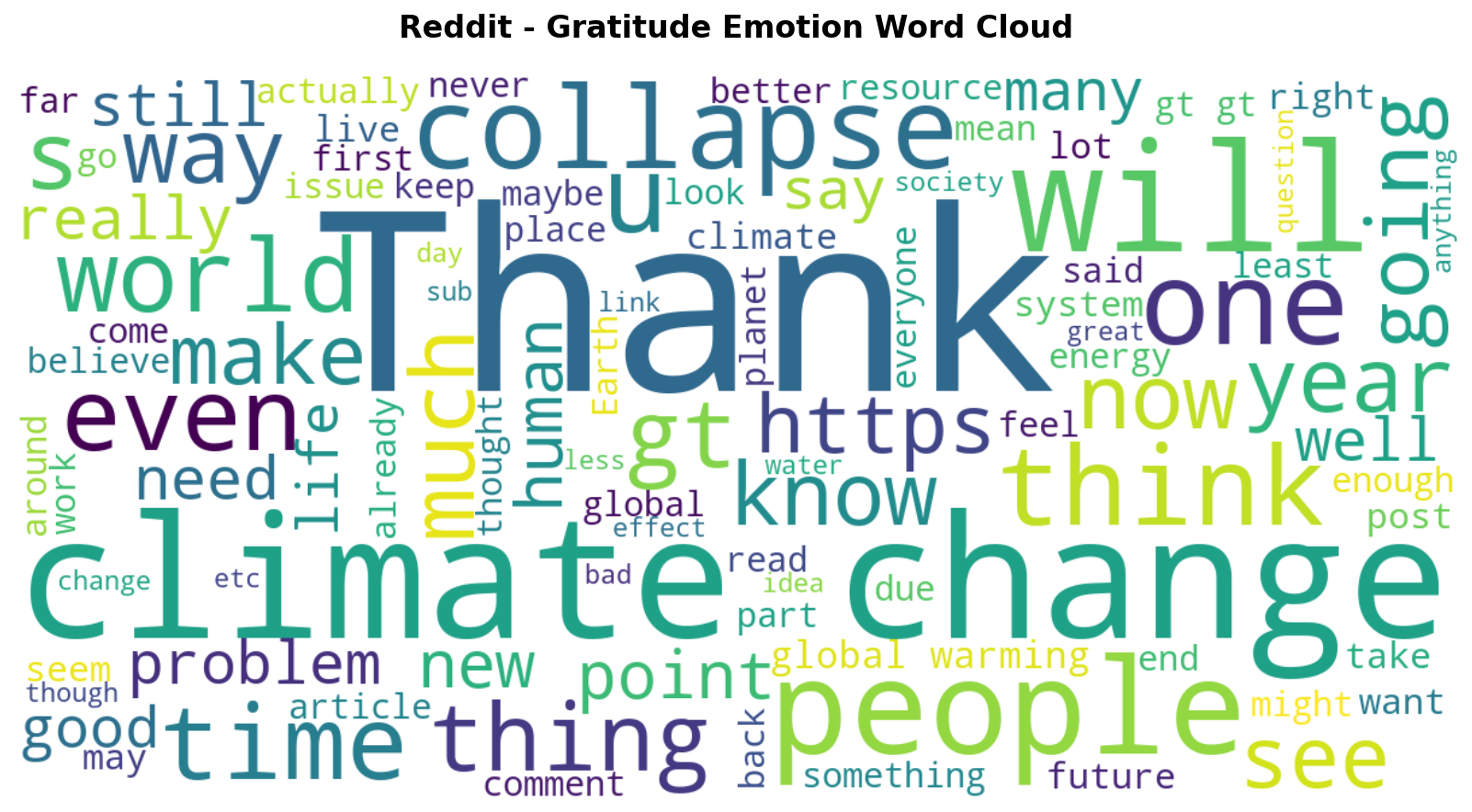}
    \caption{Gratitude}
\end{subfigure}
\hfill
\begin{subfigure}{0.48\textwidth}
    \centering
    \includegraphics[width=\linewidth]{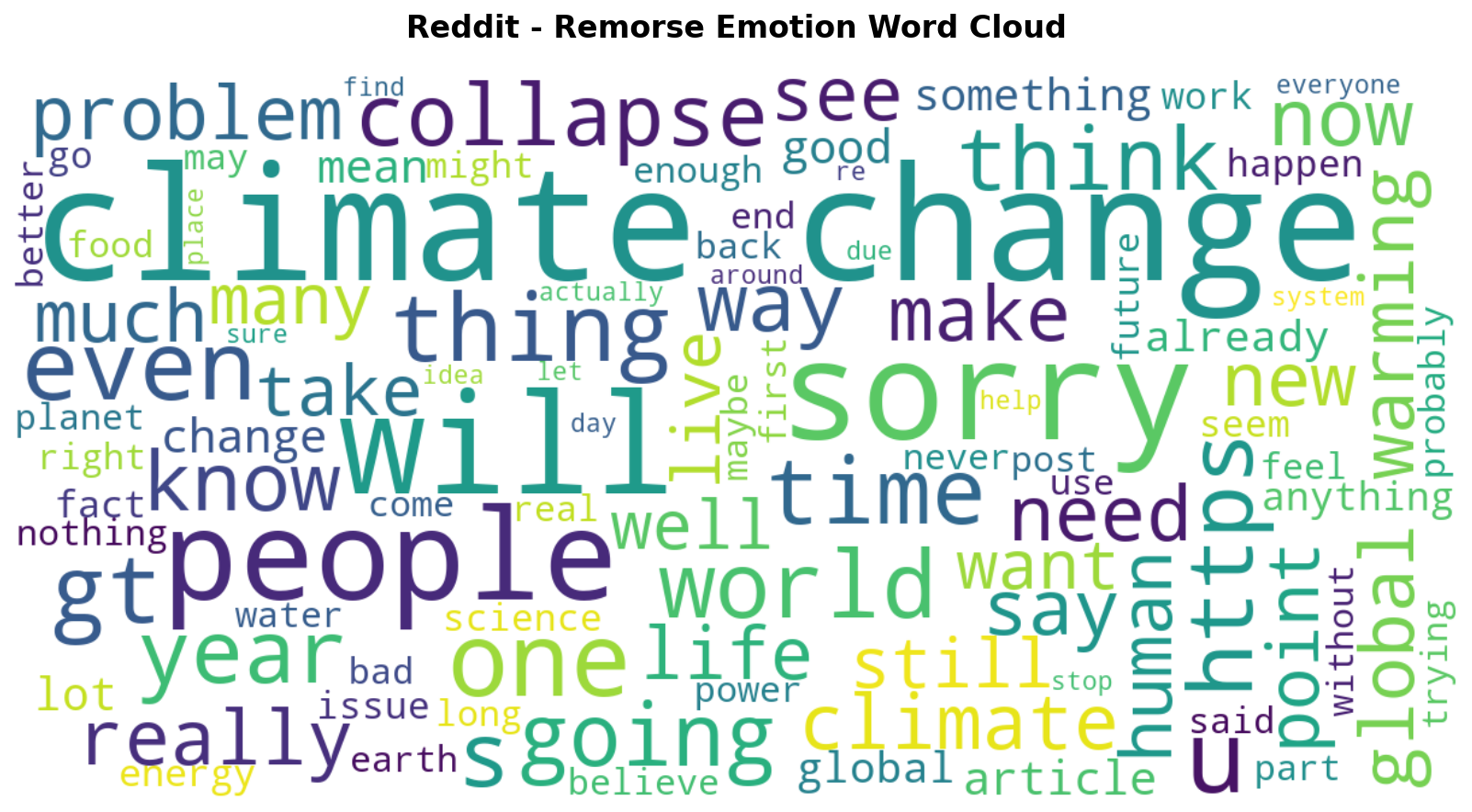}
    \caption{Remorse}
\end{subfigure}

\vspace{0.1cm}

\begin{subfigure}{0.48\textwidth}
    \centering
    \includegraphics[width=\linewidth]{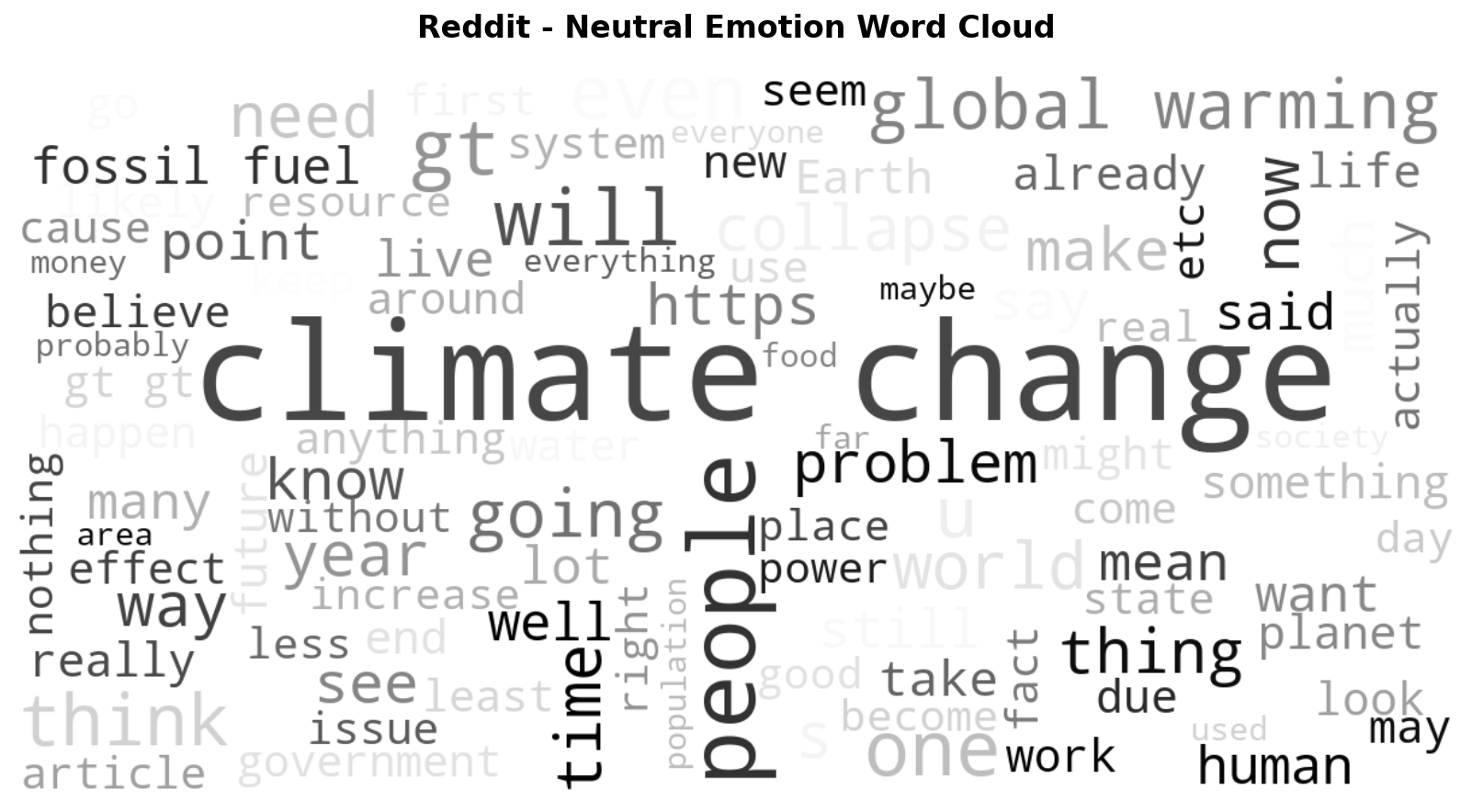}
    \caption{Neutral}
\end{subfigure}
\hfill
\begin{subfigure}{0.48\textwidth}
    \centering
    \includegraphics[width=\linewidth]{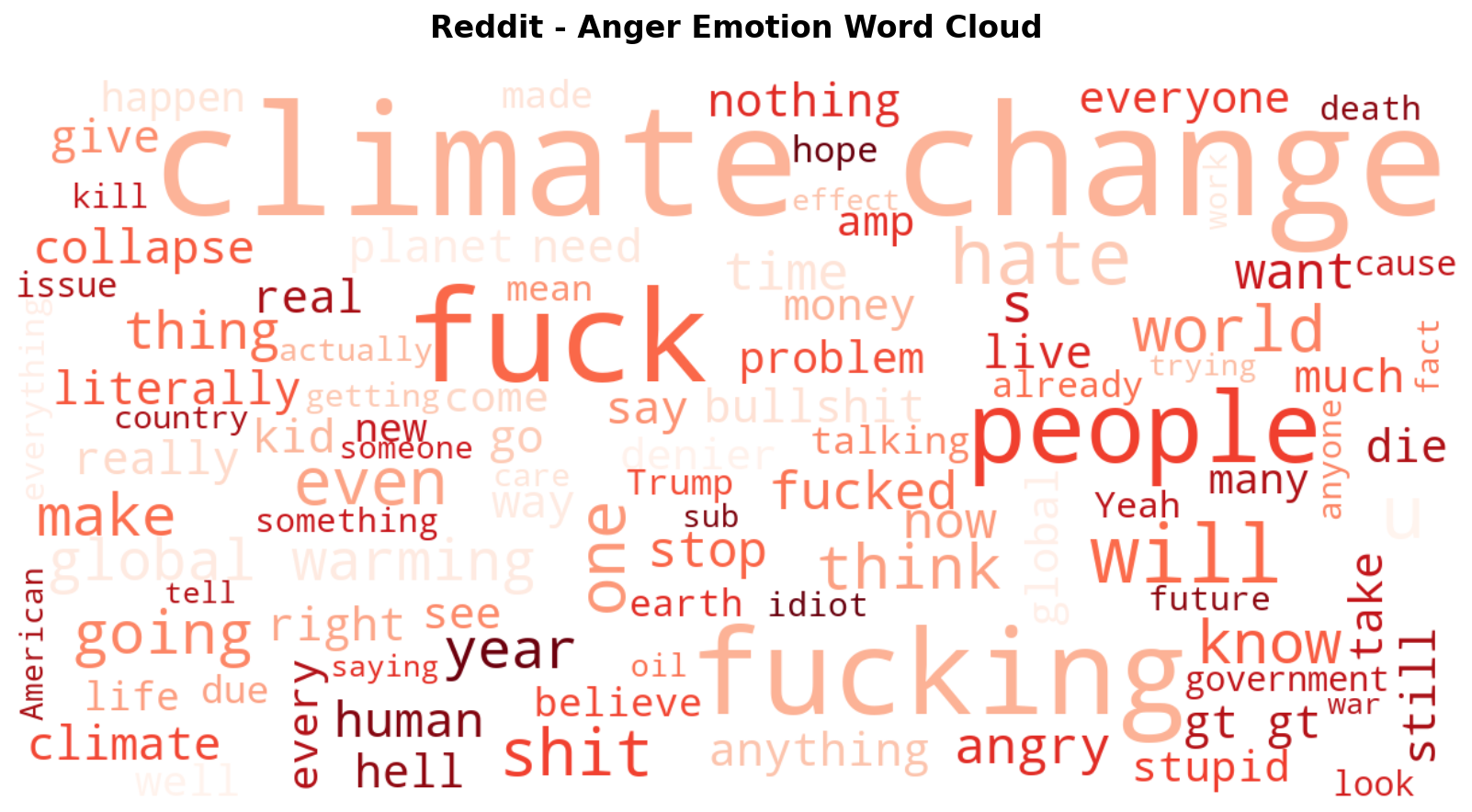}
    \caption{Anger}
\end{subfigure}

\caption{Word clouds illustrating the most prominent vocabulary associated with each predicted emotion across the complete Reddit dataset. Word size reflects relative frequency within each emotion category.}
\label{fig:wordclouds-page1}
\end{figure}

Taken together, these findings suggest that climate-related discourse contains both risk and protective factors for psychological well-being. Fear, sadness, and disappointment may contribute to anxiety and emotional distress, while gratitude, support, and collective action narratives may foster resilience and hope.

\subsection{Overall Implications}

The results suggest that climate anxiety is both an emotional and social phenomenon. Climate-related concerns are not expressed solely through fear and sadness but are shaped through interaction, discussion, and community response.

The findings indicate that online communities can play a dual role: they may amplify awareness of climate-related risks while simultaneously providing emotional support and opportunities for collective engagement. For policymakers, educators, and mental health professionals, these results highlight the importance of interventions that move beyond reducing anxiety alone and instead focus on strengthening resilience, social support networks, and meaningful pathways for climate action.

Such approaches may help transform climate anxiety from a source of distress into a catalyst for adaptive behaviour, community engagement, and long-term climate resilience.

\section{Limitations \& Future Work}\label{limitations}

This study is limited to English-language Reddit data, which reflects the communication patterns of a specific subset of social media users rather than the broader youth population. Reddit emphasises longer-form, discussion-oriented interactions, whereas many younger users engage primarily through platforms such as TikTok, Instagram, or Discord. Consequently, the findings should be interpreted as representative of climate-related discussions within Reddit communities rather than of all youth perspectives or online discourse.

Because this study analyses observational social media data, the identified relationships should not be interpreted as causal. The results describe patterns of online discussion rather than demonstrating causal links between climate-related content and psychological outcomes.

Topic modelling results depend on embedding quality, clustering parameters, and dimensionality reduction choices. While BERTopic enables flexible topic discovery, different parameter settings (UMAP \& HDBSCAN configurations) may produce different topic structures. As a result, identified topics should be interpreted as approximate groupings rather than definitive categories.

Emotion detection and topic modelling can struggle with sarcasm, slang, short posts, and rapidly evolving youth language, increasing the risk that emotions or themes are misclassified or oversimplified. In addition, differences between the models' training data and climate-related discussions may reduce performance when emotions are subtle or mixed. The current version of the framework also lacks formal fairness and robustness evaluations across demographic groups and platforms, meaning some observed patterns may reflect model behaviour rather than true differences in sentiment.

This analysis examines online expressions of climate-related experiences but does not capture offline behaviours such as coping strategies, help-seeking, or participation in climate action. Ethical risks remain, including the potential to overgeneralise findings beyond the sampled population or misinterpret aggregate patterns as measures of individual mental health. In addition, the absence of direct youth engagement limits the ability to validate interpretations. Although temporal analyses were performed using dynamic topic modelling, the findings remain bounded by the time window represented in the dataset. Longer-term shifts in climate anxiety may differ as public discourse and external events continue to evolve.

This study does not include manual annotation or ground-truth validation of topics or emotion labels. As a result, the accuracy of model outputs is not formally evaluated, and some identified patterns may reflect model bias or error rather than true underlying signals. Manual annotation and benchmark evaluation will be incorporated in future versions of the framework to better quantify model performance.

\textbf{Future Work.} Future work will focus on jointly modelling topics and emotions to better understand how climate-related themes are associated with emotional responses and how these relationships evolve over time. We also plan to develop a validated youth-focused taxonomy for defining and labelling climate anxiety, and work on the identification of youth-written content. Additional work will evaluate model fairness and robustness while possibly expanding the framework to multilingual and youth-relevant platforms. Future studies will also strengthen ethical governance through youth engagement, and enhance the ClimateLens web application and accompanying documentation to improve accessibility, reproducibility, and interactivity.

\section{Conclusion}

This study presents ClimateLens' beginning framework for analysing climate-related discussions on Reddit using topic modelling and emotion classification. By combining thematic and emotional analyses, we identify recurring discussion patterns, examine how emotions vary across communities and interactions, and explore how climate-related discourse evolves over time.

Our findings suggest that climate-related conversations extend beyond expressions of fear or sadness. Topic modelling revealed a clear distinction between awareness-oriented and action-oriented discussions, while emotion analysis demonstrated that community interactions often reshape the emotional tone of conversations by introducing support, encouragement, and alternative perspectives. Together, these results highlight that climate anxiety is both an emotional and social phenomenon, influenced not only by the topics being discussed but also by how individuals engage with one another online.

Beyond characterising climate anxiety, this work demonstrates how natural language processing can support large-scale exploration of complex social and psychological issues. As an initial preprint, this work establishes a foundation for future research on climate anxiety in digital spaces. Future iterations will expand model validation, improve youth-centred analyses, incorporate additional platforms and languages, and strengthen fairness, robustness, and ethical evaluation. We hope ClimateLens serves as both a reproducible research framework and a practical tool for understanding how online communities experience, discuss, and respond to climate-related anxiety.

\section{Acknowledgments}
The authors would like to thank individuals and organisations who provided valuable feedback, technical guidance, or intellectual support but did not meet authorship criteria (contributing significantly to a section of the preprint). We thank members of Sprout Climate Association (formerly known as Climate Resilient Communities), particularly,

\begin{itemize}
  \item Helena (Qiaoshan) Yu and Katelyn Macdonald for strategic guidance and review of this preprint
  \item Amanda Easson for early project management
  \item Maryam Tavakoli for the conceptualisation and implementation of topic modelling and emotion analysis
  \item Paras Jamil for Azure cloud configuration and technical guidance
  \item Ali Hürriyetoğlu for reviewing the v1 manuscript and providing feedback on presentation and wording
  \item Jorge Rivera for early web application development
  \item Luis Ticas and Jordan Lenard David for cloud and technical support
\end{itemize}

We also acknowledge the open-source contributors to BERTopic, Hugging Face Transformers, and related natural language processing libraries that enabled this work.

All acknowledgments are made with permission, and contributions are recognised respectfully and transparently.

\section{Declarations}

\textbf{Ethics Approval}.
This study analyses publicly available social media content from Reddit. All data were collected in accordance with the platforms’ terms of service and developer policies. No private, deleted, or access-restricted content was used. Usernames, profile information, and direct identifiers were removed prior to analysis, and results are reported only in aggregate form.

Because the study relies exclusively on publicly available, non-interactive data and does not involve intervention, experimentation, or collection of private information, Institutional Review Board (IRB) review was not required under U.S. federal regulations (45 CFR 46). Nonetheless, ethical risks such as misinterpretation, stigmatisation, or misuse of findings were carefully considered, and safeguards were implemented to minimise harm.

\textbf{Data Availability}.
Due to platform policies, raw Reddit data cannot be redistributed. However, all data collection procedures, preprocessing steps, and filtering criteria are fully documented to enable replication.

Researchers with appropriate API access can reproduce the dataset by following the instructions provided in the project repository.

\textbf{Code Availability}.
All code used for data collection, preprocessing, topic modelling, emotion classification, and visualisation is publicly available and fully documented, including setup and execution instructions in a \texttt{README}, environment specifications and dependency lists, scripts for preprocessing, topic modelling, emotion classification, and visualisation, Azure virtual machine setup resources, and example configuration files to support reproducibility.

\begin{center}
\url{https://github.com/Climate-Resilient-Communities/ClimateLens/tree/main}
\end{center}

\textbf{Competing Interests}. The authors declare that they have no competing interests.

\textbf{Author Contributions}. Karim contributed to the methodology, technical development (topic modelling and emotion analysis), and writing (review and editing). Zainab contributed to product management, the literature review, and writing (review and editing). Ardavan contributed to technical development (topic modelling, temporal analysis, and visualisation) and writing (methodology). Vikrant contributed to writing (results analysis and discussion) and product design. Bobbie contributed to writing (limitations and future work) and web app development. All authors, along with Helena Yu and Katelyn Macdonald, reviewed and approved the final manuscript.

\section{Appendices and Supplementary Materials}\label{appendix}

To support transparency, reproducibility, and further exploration of the ClimateLens framework, supplementary materials are provided in the supplementary source archive accompanying this arXiv submission. These materials include:

\begin{itemize}
    \item Data schema describing the structure of the processed datasets, preprocessing pipeline, and summary statistics.
    \item Emotion label distributions and emotion-specific word clouds for all analysed Reddit communities, including those presented in Section~\ref{emotion-and-sentiment-patterns} and Section~\ref{word-level-emotional-signals}, respectively. Moreover, emotion time-series plots are included.
    \item Full topic representations and dynamic topic evolution outputs for all discovered topics.
    \item Code snippets illustrating the core components of the pipeline. Complete source code available in repository.
    \item Documentation comparing the embedding models evaluated for topic modelling.
    \item Documentation comparing transformer-based emotion classification models for social media text.
    \item The custom stopword and preserved-word lists used during preprocessing.
    \item Parameters used throughout topic modelling.
\end{itemize}

Some supplementary materials include resources prepared for future versions of the ClimateLens framework that analyse both Reddit and Twitter/X data. The analyses presented in this preprint are based exclusively on Reddit, as described in Section~\ref{results and discussion}.



\bibliographystyle{plain} 
\bibliography{references}

\end{document}